\documentclass[aps,preprint,showpacs,floatfix]{revtex4}
\usepackage{epsfig}
\usepackage{amsmath,amsfonts,bm}
\begin{document}

\count255=\time\divide\count255 by 60 \xdef\hourmin{\number\count255}
  \multiply\count255 by-60\advance\count255 by\time
 \xdef\hourmin{\hourmin:\ifnum\count255<10 0\fi\the\count255}

\newcommand\<{\langle}
\renewcommand\>{\rangle}
\renewcommand\d{\partial}
\newcommand\LambdaQCD{\Lambda_{\textrm{QCD}}}
\newcommand\tr{\mathop{\mathrm{Tr}}}
\newcommand\+{\dagger}
\newcommand\g{g_5}

\newcommand{\xbf}[1]{\mbox{\boldmath $ #1 $}}

\title{Pion Form Factor in Improved Holographic QCD Backgrounds}

\author{Herry J. Kwee}
\email{Herry.Kwee@asu.edu}

\author{Richard F. Lebed}
\email{Richard.Lebed@asu.edu}

\affiliation{Department of Physics, Arizona State University, Tempe,
AZ 85287-1504}

\date{December 2007}

\begin{abstract}
We extend a recent numerical calculation of the pion electromagnetic
form factor $F_\pi (Q^2)$ in holographic QCD to study two important
issues regarding the behavior of fields in the bulk.  First, we show
that using a chiral symmetry-breaking field formally satisfying the
boundary conditions of the ``soft-wall'' model changes numerical
results very little from the earlier calculation that ignores these
constraints.  Second, we use a background field that interpolates
between ``hard-wall'' and ``soft-wall'' models to obtain an improved
model that reproduces the desirable phenomenological features of both.
In all cases, $F_\pi$ for large $Q^2$ is shallower than data, an
effect that can be cured by relaxing the fit to one of the static
observables, particularly the decay constant $f_\pi$.
\end{abstract}


\maketitle

\section{Introduction} \label{intro}

One of the most interesting developments in QCD in recent years has
been the application to hadronic physics of the gauge/gravity
correspondence~\cite{AdSCFT} between strongly-coupled gauge theories
and weakly-coupled gravity on curved spacetime backgrounds.  This
correspondence is most firmly established between ${\cal N} \!  = \!
4$ supersymmetric Yang-Mills theories (which are conformal field
theories [CFT]) and a 5-dimensional anti-de~Sitter (AdS) gravity
background---hence the moniker AdS/CFT\@.  However, one may conjecture
that a similar connection holds for any strongly-coupled gauge theory
possessing an approximate conformal symmetry, a property that holds
for QCD in its high-energy limit.  The terms ``AdS/QCD'' or
``holographic QCD'' refer to studies based upon the premise that QCD
belongs to this class and possesses a suitable gravity dual in 5D
space.  Properties such as confinement and chiral symmetry breaking
arise in many models built in this way; indeed, in the ``top-down''
approach one starts with a string theory and chooses a gravity
background that reproduces such basic QCD features.  Much more
amenable to phenomenological analysis is the ``bottom-up'' approach,
in which one begins with the QCD observables and determines what
gravity backgrounds reproduce them most successfully.

The fields of our 4D universe occupy only one surface of the 5D AdS
space, whose metric is given by
\begin{equation}
ds^2 = g^{\vphantom\dagger}_{MN} \, dx^M dx^N = \frac{1}{z^2}
(\eta_{\mu \nu} dx^\mu dx^\nu - dz^2) \, . \label{metric}
\end{equation}
Here the full nontrivial 5D metric $g^{\vphantom\dagger}_{MN}$
obtained from Eq.~(\ref{metric}) is distinguished from the 4D
Minkowski metric $\eta_{\mu \nu} \! = \! \rm{diag} (+,-,-,-)$.  The
ultraviolet (UV) limit of QCD is represented by fields living near the
AdS singularity $z \!  = \! \epsilon \! \to \! 0$ (the ``UV brane''),
suggesting the association of the extra ``bulk'' coordinate $z$ with
inverse momentum scales: $Q \! \sim \! 1/z$.  The gauge/gravity
correspondence states that every CFT operator ${\cal O}(x)$ is
associated with a bulk field $\Psi (x,z)$ of given quantum numbers
uniquely determined by its value $\Psi (x, \epsilon)$ on the UV brane,
which explains the origin of the usage ``holographic''.  Additionally,
the global QCD symmetry of isospin associated with the two light quark
flavors is promoted to a gauged SU(2) symmetry respected by the bulk
fields.

From the point of view of hadronic physics, the original ${\cal N} \!
= \! 4$ super-Yang Mills theory is inadequate due to being exactly
conformal and therefore lacking asymptotic $S$-matrix particle states.
In AdS/QCD one breaks the conformal symmetry (and introduces a mass
scale) by impeding the ability of the fields $\Psi (x,z)$ to penetrate
deeply into the bulk, leading to an explanation of confinement and
more generally constraining the model's infrared (IR) behavior.  The
most straightforward realization uses a hard cutoff at a particular
value $z \! = \!  z_0 \! \sim \! 1/\Lambda_{\rm QCD}$ (the ``IR
brane'') at which point appropriate boundary conditions are imposed
upon $\Psi (x,z)$, thus defining the so-called ``hard-wall''
model~\cite{PS}.  However, despite its economy, the hard-wall model
leads to results incompatible with the expectations of linear
confinement: For any model in which $\Psi (x,z)$ penetrates to a
limited depth into the bulk, the eigenmodes of $\Psi$ (with mass
eigenvalues $m_n$) represent hadronic states all carrying the same
quantum numbers, producing the AdS/QCD version of Regge trajectories.
However, QCD with linear confinement is expected~\cite{Shifman:2005zn}
to follow the trajectory $m_n^2 \! \sim \!  n$, while hard-wall models
predict $m_n^2 \! \sim \! n^2$.  To repair this shortcoming requires
the introduction of a ``soft-wall'' gravity background with an
exponential decrease $\sim e^{-\kappa^2 z^2}$ in the action for large
$z$~\cite{KKSS}, in which case $\kappa$ serves the role of
$\Lambda_{\rm QCD}$.

Studies of hadronic properties in AdS/QCD have been very popular in
recent years; to name but a few that share the same spirit as this
work are examinations of hadronic
spectra~\cite{EKSS,KKSS,BoschiFilho:2002vd,de
Teramond:2005su,Evans:2006ea,Hong:2006ta,Colangelo:2007pt,
Forkel:2007cm}, the couplings of hadrons in the presence of chiral
symmetry breaking~\cite{EKSS,Da
Rold:2005zs,Hirn:2005nr,Ghoroku:2005vt,HuangZuo}, the quark-quark
potential~\cite{heavyquark}, 4-quark operators~\cite{Hambye:2005up},
and hadronic form
factors~\cite{Hong:2004sa,GR1,GR2,Radyushkin:2006iz,BdT,GR3}.  In this
paper we continue a study of the pion electromagnetic form factor
$F_\pi$ initiated by the present authors in Ref.~\cite{KL}, using the
treatment of chiral symmetry-breaking effects (parametrized by a light
quark mass $m_q$ and chiral condensate $\sigma$) explicitly
incorporated into the Lagrangian, in the manner developed by the model
of Ref.~\cite{EKSS}.  $F_\pi$ has also been considered in two other
recent papers~\cite{BdT,GR3}; Ref.~\cite{BdT} uses a model that does
not incorporate chiral symmetry breaking explicitly, while
Ref.~\cite{GR3} also uses the formalism of Ref.~\cite{EKSS} and has a
considerable overlap~\footnote{However, we do not entirely agree with
the normalization of their parameters $f_\pi$ and $g_5$.  If we use
their values, the numerical agreement is perfect.}  with the hard-wall
calculations of Ref.~\cite{KL}; indeed, it also proves a number of
analytical results that hold in the limit $m_q \! = \! 0$.  However,
Ref.~\cite{GR3} does not present calculations in the soft-wall model
due to objections that we address below.

While results for most hadronic quantities tend to agree surprisingly
well with the results of the hard-wall model (and generally somewhat
less well for the soft-wall model~\cite{GR2,KL}), observables in the
vector meson sector tend to depend somewhat less sensitively on the
precise nature of the IR boundary condition.  However, the
axial-vector sector and especially the pion depend upon chiral
symmetry-breaking in a much more direct fashion~\cite{KKSS}; exploring
this sensitivity was an original motivation of the studies in
Refs.~\cite{GR3,KL}.

Our previous work~\cite{KL} numerically calculated $F_\pi (Q^2)$ for
spacelike values $Q^2 \! \ge \! 0$ of momentum transfer, as well as
couplings $g_{\rho^{(n)} \pi \pi}$ obtained from the timelike region,
in both the hard-wall and soft-wall models.  Our results showed that
both models predict $Q^2$ dependence for $F_\pi (Q^2)$ that tends to
be too shallow for both models (in particular, predicting a value too
small for the charge radius $\langle r_\pi^2 \rangle$), but
significantly worse for the soft-wall case.  Better agreement with
$F_\pi (Q^2)$ data could be achieved only by loosening the fit to
other observables, especially by decreasing the decay constant
$f_\pi$.

However, a small swindle is introduced in applying the methods of
Ref.~\cite{EKSS} to the soft-wall model~\cite{KKSS}: The background
scalar field $X(z)$ whose vacuum expectation value $X_0(z)$ provides
$m_q$ and $\sigma$ via its small-$z$ asymptotic value ($2X_0(z) \! =
\! m_q \! z \! + \!  \sigma z^3)$ satisfies in general a 2nd-order
differential equation whose solutions are Kummer functions, only one
of which [$zU(\frac 1 2, 0 ,\kappa^2 z^2)$, to be precise] satisfies
the necessary finiteness boundary condition of the soft-wall model as
$z \! \to \! \infty$.  But this particular function carries a unique
proportionality between the $z$ and $z^3$ terms of its Taylor series,
implying an unphysical fixed ratio between $m_q$ and $\sigma$.  Such
an obstacle is easily overcome by including higher-order terms in
$X(z)$ in the potential that do not greatly modify the small-$z$
behavior; however, from the operational point of view, in
Ref.~\cite{KL} we simply observed that such large-$z$ modifications
are numerically heavily suppressed by the background factor
$e^{-\kappa^2 z^2}$.  One of the goals of the current work is to
present a calculation using a field $X_0(z)$ that satisfies both
constraints, finite for large $z$ and allowing independent $m_q$ and
$\sigma$ values.  Numerous functional forms for the field $X_0(z)$
satisfy this requirement, but for sake of definiteness we choose
\begin{equation}
2X_0(z)={\left(m_q \, z + \sigma \, z^3 \right)} \left[1-{\rm exp}
\left(-\frac{A_c}{\kappa^4\,z^4}\right)\right] + B_c \, {\rm
exp}\left(-\frac{3}{4\,\kappa^2\,z^2}\right). \label{eq:mod_vev}
\end{equation}
We discuss the motivation for this particular choice in
Sec.~\ref{results}, but point out that numerical simulations for
several other similar forms do not significantly alter our conclusion:
Using a functional form satisfying all required asymptotic
behaviors does not significantly change the naive soft-wall fit of
Ref.~\cite{KL}.

The other goal of this paper is to develop a model that combines the
numerical successes of the hard-wall model with the Regge trajectories
predicted by the soft-wall model.  In the closing statements of
Ref.~\cite{KL} we presented a background field with precisely these
properties.  Inspired by the old Saxon-Woods generalization of the
hard-cutoff nuclear density model, we proposed
\begin{equation} \label{SW}
e^{-\Phi(z)} = \frac{e^{\lambda^2 z_0^2} - 1}{e^{\lambda^2 z_0^2} +
e^{\lambda^2 z^2} - 2} \, ,
\end{equation}
which has a drop-off at $z \! = \! z_0$ (and equals $\frac 1 2$
there), but decreases as $e^{-\lambda^2 z^2}$ for large $z$.

Our results indicate that neither of these modifications substantially
alter the outcome of Ref.~\cite{KL}.  The soft-wall model test
function $X_0(z)$ of Eq.~(\ref{eq:mod_vev}), although formally
possessing the correct asymptotic behaviors, gives an optimal fit to
$F_\pi (Q^2)$ no better than that in Ref.~\cite{KL}.  Similarly, as
expected, the adoption of Eq.~(\ref{SW}) for the background field
interpolates results between hard- and soft-wall models.  This latter
result may be somewhat surprising since Eq.~(\ref{SW}) contains two
independent mass parameters ($1/z_0$ and $\lambda$) rather than the
one of the strict hard- or soft-wall models.  Since our previous
work~\cite{KL} showed that both hard- and soft-wall model predictions
for $F_\pi (Q^2)$ can be improved significantly by loosening the fit
to just one observable, one might expect that Eq.~(\ref{SW}) with its
extra fit parameter allows sufficient latitude not only to give both
hard- and soft-wall behaviors, but also to improve the global
numerical fit to $F_\pi (Q^2)$ and other low-energy observables.  As
seen below, the former goal is achieved but the latter is not.
Taking these results together, it appears that achieving a better
simultaneous fit to $F_\pi (Q^2)$, $f_\pi$, $m_\pi$, $m_\rho$,
$f_\rho$, and other low-energy observables requires a more general
treatment of chiral symmetry breaking than the one developed in
Ref.~\cite{EKSS}.

This paper is organized as follows: In Sec.~\ref{formalism} we recount
the formalism of Ref.~\cite{EKSS} relevant to our calculations and
give expressions for $F_\pi (Q^2)$ and couplings in terms of AdS/QCD
mode wave functions.  Section~\ref{results} presents the results of
numerical simulations of $F_\pi (Q^2)$ and other low-energy hadronic
observables, and Sec.~\ref{concl} summarizes our results and
concludes. Details of the numerical procedure are discussed in
the Appendix.

\section{Formalism} \label{formalism}

The content of this section is nearly identical to that in
Ref.~\cite{KL}, but is included for completeness.  The full 5D
action~\cite{EKSS} used in this work reads
\begin{equation}\label{5DL}
S = \int\! d^{\, 5} \! x \: e^{-\Phi(z)} \, \sqrt{g}\, \tr \left\{
|DX|^2 + 3 |X|^2 - \frac1{4\g^2} (F_L^2 + F_R^2) \right\} \, ,
\end{equation}
where $g \! \equiv \! | \det g^{\vphantom\dagger}_{MN} |$ is obtained
from the metric in Eq.~(\ref{metric}), and $e^{-\Phi(z)}$ represents a
background dilaton coupling.  The original holographic QCD calculation
in Ref.~\cite{EKSS} uses the hard-wall (step function) background
$e^{-\Phi(z)} \!  = \! H (z_0 \! - \! z)$, while Ref.~\cite{KKSS}
defines the soft-wall model by $e^{-\Phi(z)} \! = \! e^{-\kappa^2
z^2}$, which as noted above, reproduces the traditional Regge
trajectory behavior for the mesons.  In this paper we also consider
the Saxon-Woods form for $e^{-\Phi(z)}$ defined in Eq.~(\ref{SW}).
The scalar field $X(z)$ [actually $(2/z) X(z)$] is the holographic
partner of the quark condensate $\bar q^{\vphantom\dagger}_R
q^{\vphantom\dagger}_L$, and its vacuum expectation value $X_0 (z) \!
\equiv \! \frac 1 2 v(z) \! = \! \frac 1 2 m_q z \! + \! \frac 1 2
\sigma z^3$ gives rise to explicit and spontaneous chiral symmetry
breaking.  This expression for $X_0 (z)$ is exact in the hard-wall
model, while it holds only for small $z$ in the soft-wall model.

The chiral gauge fields $A^a_{L,R}$ (holographic partners to the
bilinears $\bar q^{\vphantom\dagger}_{L,R} \gamma^\mu t^a
q^{\vphantom\dagger}_{L,R}$) enter through $D^M X \! \equiv \! \d^M \!
X \! - \! iA_L^M X \! + \!  iX A_R^M$, $A_{L,R}^M \! \equiv \!
A_{L,R}^{M \, a} \, t^a$, and $F_{L,R}^{MN} \! \equiv \d^M \!
A_{L,R}^N \! - \! \d^N \! A_{L,R}^M \! - \! i[A_{L,R}^M , A_{L,R}^N]$.
The polar $V$ and axial $A$ gauge fields are $V^M \! \equiv \! \frac 1
2 (A_L^M \! + \!  A_R^M)$ and $A^M \! \equiv \! \frac 1 2 (A_L^M \! -
\! A_R^M)$, in terms of which $D^M X \! = \! \d^M \! X \! - \!  i
[V^M \! , X] \! - \! i \{A^M , X \}$, $F_V^{MN} \! \equiv \! \d^M \!
V^N \! - \! \d^N \! V^M \! - \! i \left( [V^M \! , V^N] \!  + \! [A^M
\! , A^N] \right)$, $F_A^{MN} \! \equiv \! \d^M \! A^N \! - \! \d^N \!
A^M \! - \! i \left( [V^M \! , A^N] \! + \! [A^M \! , V^N] \right)$,
and
\begin{equation}\label{5DL_V_A}
S = \int\! d^{\, 5} \! x \, e^{-\Phi(z)} \! \sqrt{g} \, \tr \left\{
|DX|^2 + 3 |X|^2 - \frac{1}{2\g^2} (F_V^2 + F_A^2) \right\} \, .
\end{equation}
We use the axial-like gauge $V_z (x,z) \!  = \! 0$, $A_z (x,z) \! = \!
0$, so that the associated sources may be expressed as divergences
over just the usual four spacetime dimensions, $\d^\mu V_\mu \! = \!
0$ (since isospin is conserved) and $\d^\mu \! A_\mu$.  $A_\mu$ can be
further decomposed into a transverse (divergenceless) piece $A_{\mu \,
\perp}$ and a longitudinal piece $\varphi$: $A_\mu \!  = \! A_{\mu \,
\perp} + \d_\mu \varphi$.  The pion field $\pi^a$ appears through $X
\! = \! X_0 \exp (2i \pi^a t^a)$; $\pi^a$ is dimensionless and related
to the canonically-normalized pion field $\tilde \pi^a$ of chiral
Lagrangians via $\pi^a \! = \! \tilde \pi^a / f_\pi$, with $f_\pi \! =
\! 93$~MeV.

The equations of motion obtained from Eq.~(\ref{5DL_V_A}) for the
fields $\Psi (q,z)$ ({\it i.e.}, all except for $X_0$ are Fourier
transformed with respect to the 4D coordinates $x$) read
\newcommand{\AT}{A^a_\mu}
\begin{equation}\label{eqVAdS}
\d_z\left(\frac{e^{-\Phi(z)}}{z} \, \d_z V_\mu^a \right)
 + \frac{q^2 e^{-\Phi(z)}}{z} V_\mu^a = 0 \, ,
\end{equation}
\begin{equation}
  \left[ \d_z\left(\frac{e^{-\Phi(z)}}{z} \, \d_z \AT \right) +
\frac{q^2 e^{-\Phi(z)}}{z}
  \AT - \frac{\g^2 \, v(z)^2 e^{-\Phi(z)}}{z^3} \AT\right]_\perp =0 \,
  , \label{AT}
\end{equation}
\begin{equation}
  \d_z\left(\frac{e^{-\Phi(z)}}{z} \, \d_z \varphi^a \right)
+\frac{\g^2 \, v(z)^2 e^{-\Phi(z)}}{z^3} (\pi^a-\varphi^a) = 0 \, ,
\label{AL}
\end{equation}
\begin{equation}
  -q^2\d_z\varphi^a+\frac{\g^2 \, v(z)^2}{z^2} \, \d_z \pi^a =0 \, ,
\label{Az}
\end{equation}
\begin{equation}
\d_z \left( \frac{e^{-\Phi(z)}}{z^3} \d_z X_0 \right) +
\frac{3e^{-\Phi(z)}}{z^5} X_0 = 0 \, .
\label{Xeqn}
\end{equation}
The gauge/gravity correspondence itself provides the first prediction,
the value of the 5D gauge coupling $g_5$~\cite{EKSS}.  Substituting
Eq.~(\ref{eqVAdS}) into the $F_V^2$ portion of Eq.~(\ref{5DL_V_A})
leaves only the boundary term for the action:
\begin{equation}
S = -\frac{1}{2\g^2} \int\! d^4x \, \left. \frac{e^{-\Phi(z)}}{z}
V_\mu^a \d_z V^{\mu a} \right|_{z=\epsilon} \, .
\end{equation}
The vector field $V_\mu^a (q,z)$ can be resolved as $V_\mu^a (q,z) =
V(q,z) \tilde V_\mu^a (q)$, where $\tilde V_\mu^a (q)$ is the Fourier
transform of the vector current source $J_\mu^a \! = \! \bar q
\gamma_\mu t^a q$ at the UV boundary $z \! = \! \epsilon$, and
$V(q,z)$ (the ``bulk-to-boundary propagator'') is normalized to
$V(q,\epsilon)=1$.  Due to the isospin conservation constraint $q_\mu
V^\mu \! = \! 0$, one may replace $\tilde V_\mu^a \tilde V^{\mu \, a}$
with $\tilde V_\mu^a \tilde V_\nu^b \Pi^{\mu \nu} \delta^{ab}$ and
$\Pi^{\mu \nu} \! \equiv \! \eta^{\mu \nu} \! - \! q^\mu q^\nu /q^2$,
and then the usual quadratic variation of the action with respect to
the source $\tilde V$ produces the vector current two-point function:
\begin{eqnarray}
  \int d^4 x \, e^{iqx} \<J_\mu^{a}(x)J_\nu^{b}(0)\> \!&=&\!
  \delta^{ab} \, \Pi_{\mu \nu} \, \Sigma_V (q^2) \, ,
  \label{eq:VV}\\
  \Sigma_V (q^2) \!&=&\! 
  \left.-\frac{e^{-\Phi(z)}}{\g^2} \frac{\d_z V(q,z)}{z}
\right|_{z=\epsilon} \, ,
\label{Vqz}
\end{eqnarray}
from which one finds, matching to the QCD result for currents $J_\mu$
normalized~\cite{GR2} according to the prescription of~\cite{EKSS},
\begin{equation} \label{g5val}
\g^2 = \frac{12 \pi^2}{N_c} \to 4\pi^2 \,.
\end{equation}

An analogous calculation in the axial sector relates the
bulk-to-boundary propagator $A(q,z)$ to the $\pi$ decay constant
$f_\pi$:
\begin{equation}\label{fpi}
f_\pi^2 = -\frac1{\g^2}\left.\frac{\d_z
A(0,z)}{z}\right|_{z=\epsilon} \, ,
\end{equation}
from which one sees that the relative choices of normalization for
$f_\pi$ and $g_5$ are correlated.

The normalizable eigenstates of Eqs.~(\ref{eqVAdS})--(\ref{Az})
correspond to towers of hadrons of the same quantum numbers as the
parent fields.  Since large $N_c$ is intrinsic to this procedure, the
mesons have narrow widths and the spectral decompositions of
self-energy functions such as $\Sigma_V$ are sums over poles:
\begin{equation} \label{spectral}
\Sigma_V (q^2) = \sum_{n=0}^\infty \frac{f_n^2}{q^2 - M_n^2} \, ,
\end{equation}
where $M_n$ are the mass eigenvalues and $f_n$ are the decay constants
of vector modes $\psi_n (z)$ normalized according to
\begin{equation}
\int dz \frac{e^{-\Phi(z)}}{z} \, \psi_m (z) \psi_n (z) = \delta_{mn}
\, . \label{psi_norm}
\end{equation}

In order to obtain the pion form factor $F_\pi$, one expands the
action Eq.~(\ref{5DL_V_A}) out to cubic order in the fields.  Since
the pion field is related [Eq.~(\ref{Az})] to the longitudinal mode
$\d^\mu \varphi$ of $A^\mu$, one must identify not only $V\pi\pi$
terms, but also $V \! AA$ and $V \! A\pi$.  A straightforward
calculation (that uses the equation of motion for $V$ to simplify the
result) produces the $V\pi\pi$ terms:
\begin{align}\label{Vpipi_Action}
S_{\rm AdS}^{V\pi\pi}  &=
 \epsilon_{abc}\int d^4 x \int dz \, e^{-\Phi(z)} \,
 \left[ \frac{1}{g_5^2\,z}
  \left(\partial_z\partial^\mu\varphi^a\right) V_{\mu}^b
  \left(\partial_z\varphi^c\right) \right. \nonumber \\
 &\left. \quad +\frac{v(z)^2}{z^3}
  \left(\partial^{\mu}\pi^a- \partial^{\mu}\varphi^a\right)
  V_{\mu}^b \left(\pi^c-\varphi^c\right)
  \right] \, ,
\end{align}
where the $z$ integration range is $[0,\infty)$.
Reference~\cite{EKSS} uses this action (but only with $e^{-\Phi(z)}
\!= \! 1$) to obtain the $V \pi \pi$ couplings [Eq.~(\ref{gnpipi})
below], with the caveat that terms cubic in $F_{V,A}$ have not been
included.  In fact, we have shown~\cite{KL} that no such terms
contribute to the $V \pi \pi$ coupling.  The 3-point correlator is
obtained from a straightforward variation of Eq.~(\ref{Vpipi_Action}):
\begin{align}
 \langle J_\pi^a(p_1)J_V^{\mu,b}(q) J_\pi^c(-p_2)\rangle =
 \epsilon^{abc} F(p_1^2,p_2^2,q^2)\left(p_1+p_2\right)^\mu i (2\pi)^4
\delta^{(4)} (p_1 - p_2 + q) \ ,
\end{align}
where, again recalling the narrowness of resonances, one may express
the dynamical factor $F(p_1^2,p_2^2,q^2)$ in terms of transition form
factors:
\begin{align}
\label{Fsuv} F(p_1^2,p_2^2,q^2) = \sum_{n,k = 1}^{\infty}
 \frac{f_{n} f_{k}  F_{nk} (q^2) }{\left(p_{1}^2 -
 M^2_{n}\right)\left(p_{2}^2 - M^2_{k}\right)}  \ ,
\end{align}
and $F_{nk}(q^2)$ correspond to form factors for $n \! \to \! k$
transitions.  The pion form factor $F_\pi (q^2)$ is then defined as
the ground-to-ground pseudoscalar meson transition to the vector
current:
\begin{equation}
F_\pi(q^2) \equiv F_{11}(q^2) = \int dz \, e^{-\Phi(z)} \,
\frac{V(q,z)}{f_\pi^2} \left\{ \frac{1}{g_5^2 z}
[\partial_z\varphi(z)]^2 + \frac{v(z)^2}{z^3} \left[\pi(z) -
\varphi(z)\right]^2 \right\} \ , \label{ff} 
\end{equation}
whose normalization $F_\pi (0) \! = \! 1$ [using $V(0,z) \! = \! 1$]
is guaranteed by the canonical normalization of the pion kinetic
energy term in the 4D Lagrangian. The origin of the $z$ dependence of
Eq.~(\ref{ff}) is recognizable from Eq.~(\ref{Vpipi_Action}).  The
pion is the ground-state solution of Eqs.~(\ref{AL})--(\ref{Az}) for
$\pi^a$ subject to appropriate boundary constraints: Neumann at $z \!
= \! z_0$ for the hard-wall model, finite as $z \! \to \! \infty$ for
the soft-wall model.  The solution to Eq.~(\ref{eqVAdS}) for source
$V$ in the soft-wall model [$\Phi(z) \! = \! \kappa^2 z^2$] for
spacelike momentum transfers $q^2 \! \equiv \! -Q^2 \! < \!
0$~\cite{KL,GR2,BdT}, used in our first calculation, includes the
Kummer (confluent hypergeometric) function $U$:
\begin{equation} \label{Vsoft}
V(q,z) = \Gamma (1 + Q^2 \! /4\kappa^2) \, U[ Q^2 \! /4\kappa^2, 0,
(\kappa z)^2 ] \, .
\end{equation}
Note that for large $z$, Eq.~(\ref{Vsoft}) falls as $(z^2)^{-Q^2/4\kappa^2}$.
For our second calculation [$e^{-\Phi(z)}$ as in Eq.~(\ref{SW})],
$V(q,z)$ must be obtained numerically from Eq.~(\ref{eqVAdS}).  In
both cases the solutions satisfy the boundary conditions
$V(q,\epsilon) \! = \! 1$, $V(0,z) \! = \! 1$.  In the timelike
region, $V(q,z)$ may be expanded as
\begin{equation}
 \label{vector_expand}
 V(q,z) = -g_5 \sum_{n = 1}^{\infty} \frac{f_{n} \psi_n( z)}
 { q^2 - M^2_{n} } \, .
\end{equation}
When numerical solutions are required, $\psi_n (z)$ and $f_n$ are
obtained by solving Eq.~(\ref{eqVAdS}) at its poles and properly
normalizing using Eq.~(\ref{psi_norm}).  In either case, substituting
Eq.~(\ref{vector_expand}) into Eq.~(\ref{ff}) gives the timelike pion
form factor as a sum over vector meson poles:
\begin{equation}
 \label{pion_timelike}
 F_\pi(q^2) = -\sum_{n = 1}^{\infty} \frac{f_{n} g_{n\pi\pi}}
 {q^2 - M^2_{n}} \ ,
\end{equation}
where $g_{n\pi\pi}$ is given by
\begin{equation} \label{gnpipi}
g_{n \pi \pi} = \frac{g_5}{f_\pi^2} \int \! dz \, \psi_n (z) \,
e^{-\Phi(z)} \left\{ \frac{1}{g_5^2 z} [\d_z \varphi (z)]^2 +
\frac{v(z)^2}{z^3} \left[ \pi (z) - \varphi(z) \right]^2 \right\} \,
\ .
\end{equation}
Together, Eqs.~(\ref{ff}) and
Eqs.~(\ref{pion_timelike})--(\ref{gnpipi}), with $V$ and $\psi_n$
obtained from either Eq.~(\ref{Vsoft}) or the numerical solution to
Eq.~(\ref{eqVAdS}), provide a complete expression for $F_\pi (q^2)$
at all values of $q^2$.

\section{Results} \label{results}

For our first result, we consider the consequences of modifying the
solution for the vev of the bulk field $X(z)$, as stated in the
Introduction and repeated here for convenience:
\begin{equation}
\setcounter{equation}{2}
2X_0(z)\equiv v(z)=\left(m_q \, z + \sigma \, z^3 \right) \left[1-{\rm
exp} \left(-\frac{A_c}{\kappa^4\,z^4}\right)\right] + B_c \, {\rm
exp}\left(-\frac{3}{4\,\kappa^2\,z^2}\right)\, . 
\end{equation}
This particular expression is chosen because of the interesting
large-$z$ asymptotic form of solutions of Eq.~(\ref{Xeqn}) for
$X_0(z)$~\cite{KKSS}: In the hard-wall case the independent solutions
are precisely $z^1$ and $z^3$ (so that the asymptotic solution is also
the full solution in that case), while for the soft-wall case the
exact solutions are the Kummer functions $z^3 M (\frac 3 2, 2,
\kappa^2 z^2)$ and $z U (\frac 1 2, 0, \kappa^2 z^2)$.  Of the latter
pair only the second is finite as $z \! \to \! \infty$, and indeed
behaves as $\exp(-3/4 \kappa^2 z^2)$.  As indicated in the
Introduction, $X_0$ must have corrections arising from higher-order
terms in the potential to allow independent $m_q$ and $\sigma$
parameters at low $z$, which in turn would appear as corrections to
Eq.~(\ref{Xeqn}).  This argument originated with the original
soft-wall paper~\cite{KKSS}, and since a full explanation that
motivates our choice Eq.~(\ref{eq:mod_vev}) requires understanding the
original argument in detail, we repeat it here.  Inasmuch as
$e^{-\Phi(z)}$ vanishes as $z \! \to \! \infty$ in the soft-wall
model, the equation of motion Eq.~(\ref{Xeqn}) for $X_0$ is guaranteed
to possess a solution that tends to a constant as $z \! \to \!
\infty$.  Since the $e^{-\Phi(z)}$ factor also multiplies any
higher-order potential terms in $X$ that might be introduced into the
action Eq.~(\ref{5DL_V_A}) and hence into Eq.~(\ref{Xeqn}), the full
equation of motion continues to possess a solution that approaches a
constant as $z \! \to \! \infty$.  Moreover, for such a solution with
$X_0 (\infty) \! =$ constant, all multiple-derivative terms appearing
in Eq.~(\ref{Xeqn}) and its generalizations become negligible compared
to the lone single-derivative term contained in the first term of
Eq.~(\ref{Xeqn}).  The constant $X_0(\infty)$ therefore acts as the
sole integration constant in this effectively 1st-order differential
equation, and thus determines the (in general nonlinear) relationship
between its solution's $z^1$ and $z^3$ coefficients, namely, the ratio
of $m_q$ to $\sigma$.  By adjusting the higher-order terms in the
potential, one changes $X_0(\infty)$, thus altering the numerical
ratio between its coefficients near $z \! = \! 0$, thereby allowing
one to obtain the desired ratio of $m_q$ to $\sigma$.

With these features in mind, let us return to the motivation for
the choice Eq.~(\ref{eq:mod_vev}).  The asymptotic form
$\exp(-3/4\kappa^2 z^2)$ already appears for the explicit solution to
Eq.~(\ref{Xeqn}) and therefore is chosen for the $z \! \to \! \infty$
limit of Eq.~(\ref{eq:mod_vev}).  The functional forms $\exp( -A/z^n
)$ with $A, n \!  > \! 0$ are especially interesting because
they not only rapidly approach 1 as $z \! \to \! \infty$, but also
sluggishly depart from 0 as $z$ is increased from 0, due to their
nonanalyticity about that point.  Thus the small-$z$ asymptotic
form of Eq.~(\ref{eq:mod_vev}), $m_q z \! + \! \sigma z^3$, is
precisely as required.  These properties make Eq.~(\ref{eq:mod_vev})
an excellent choice for a test function satisfying the desired
asymptotic behaviors.  Note that we do not exhibit a specific
potential with this solution to its equation of motion, since
Eq.~(\ref{eq:mod_vev}) satisfies all the requirements placed upon it
through its solution.  Moreover, Eq.~(\ref{eq:mod_vev}) gives results
for $F_\pi (Q^2)$ and other observables in particular ranges of its
parameters $A_c$ and $B_c$ that agree numerically with the original
soft-wall model (which, recall, is the exact solution for a particular
potential), thereby providing additional ({\it a posteriori}) support
for this choice.  In particular, the precise form of $X_0$ at
intermediate values of $z$---ignored in the construction of
Eq.~(\ref{eq:mod_vev})---is apparently not important in fitting
observables.

\begin{figure}[t]
\epsfxsize 5.0 in \epsfbox{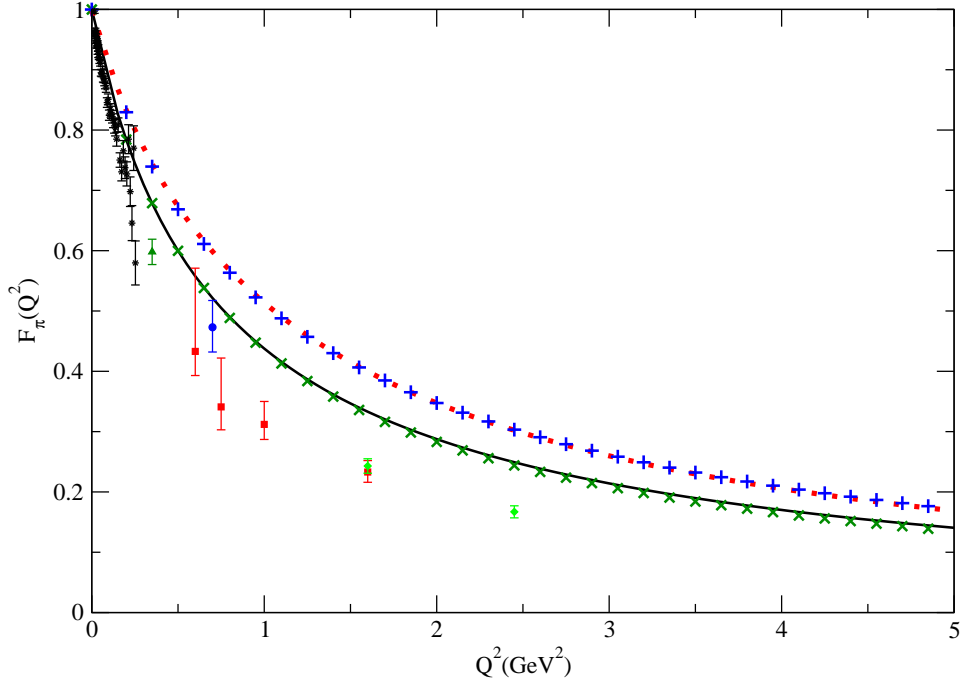}
\caption{Spacelike scaling behavior for $F_\pi(Q^2)$ as a function of
$Q^2 \! = \! -q^2$.  The continuous line is the prediction of the
original hard-wall model with $1/z_0 \! = \! 323$~MeV.  The dotted
line is the prediction of the original soft-wall model with $\kappa \!
= \! m_\rho/2$.  The crosses use $e^{-\Phi(z)}$ of Eq.~(\ref{SW}) with
$\lambda z_0 \! = \! 2.1$, and the pluses use $\lambda z_0 \! =
\! 1$.  The
stars are from a data compilation from CERN~\cite{Ame84}, the
circles are from DESY, reanalyzed by Tadevosyan {\it et
al.}~\cite{Bra77,Tadevosyan:2007yd}, the
triangle is data from DESY~\cite{Ack78}, and the
boxes~\cite{Tadevosyan:2007yd} and
diamonds~\cite{Horn:2006tm} are from Jefferson Lab.  Older data in the
range 3--10~GeV$^2$~\cite{bebek} exist but have large uncertainties
and are not presented here.}
\label{fig:formfactor}
\end{figure}

As mentioned above, the numerical change from the results of
Ref.~\cite{KL} (the original soft-wall model curve in
Fig.~\ref{fig:formfactor}) due to using Eq.~(\ref{eq:mod_vev}) turns
out to be quite minimal in the best-fit case.  For sake of
definiteness, let us choose $B_c\!=\!1$ and $A_c\!=\!1$.  We find
these values give a best fit to $F_\pi(Q^2)$ data quite close to (but
slightly worse than) that of the original soft-wall model obtained in
Ref.~\cite{KL}---an encouraging start.  Keeping $A_c\!=\!1$ and
varying $B_c$, we find that values of $B_c \!  < \! 1$ (including $B_c
\! < \! 0$) give an $F_\pi$ prediction that moves even further away
from the data ({\it i.e.}, is even more shallow) than the original
soft-wall prediction.  On the other hand, for $B_c \! > \! 1$ the
prediction moves closer to $F_\pi$ data, but still is only as good as
the original soft-wall prediction at $B_c \approx 2$.  For $B_c \!
\agt \! 2$ the prediction again moves away from the $F_\pi$ data.
Next, if we keep $B_c \!  = \! 1$ and vary $A_c$, we find that for $0
\! < \! A_c \! < \! 1$ the prediction moves away from the $F_\pi$
data, but for $A_c \!  > \! 1$ the prediction improves only
marginally.  Interestingly, letting both $B_c$ and $A_c$ vary always
appears to permit a set of parameters that fit the data [$F_\pi$ and
the static observables $m_\pi$, $m_\rho$, $m_{a_1}$, $f_\pi$,
$f_\rho^{1/2}$, $f_{a_1}^{1/2}$, and $g_{\rho \pi \pi}$] reasonably
well, but only if $B_c \! \alt\! 2$.  However, these fits are never
superior to the original soft-wall fit.  We conclude that the original
fit in Ref.~\cite{KL}, although formally violating the AdS/QCD
boundary conditions, is actually quite stable when the proper boundary
conditions are imposed.
\begin{table}[t]\vspace{-6pt}
\caption[]{Comparison of soft-wall model to modified $e^{-\Phi(z)}$
with $\lambda z_0\!=\!1$; values in MeV (except for $g_{\rho \pi
\pi}$).}
\begin{ruledtabular}
\begin{tabular}{cccc}  
Observable & Experiment & Soft-wall &$\lambda z_0 \!=\!1$\\
\hline
$m_\pi$            & 139.6$\pm0.0004$ \cite{PDG}          & 139.6      & 139.6 \\
$m_\rho$           & 775.5$\pm 0.4$   \cite{PDG}          & 777.4      & 779.2 \\
$m_{a_1}$          & 1230$\pm40$      \cite{PDG}          & 1601       & 1596  \\
$f_\pi$            & 92.4$\pm0.35$    \cite{PDG}          & 87.0       & 92.0  \\
$f_\rho^{\,1/2}$   & 346.2$\pm1.4$    \cite{Donoghue}     & 261        & 283   \\
$f_{a_1}^{\,1/2}$  & 433$\pm13$     \cite{SS,Isgur:1988vm}& 558        & 576   \\
$g_{\rho\pi\pi}$   & 6.03$\pm0.07$    \cite{PDG}          & 3.33       & 3.49  \\
\end{tabular}
\end{ruledtabular}
\label{Table1}
\end{table}

That the new fits never improve upon the original model is quite easy
to explain numerically.  First note that $V(q,z)$ in Eq.~(\ref{Vsoft})
drops from $1$ at $z\!=\!0$ to smaller values for larger $z$ with
increasing rapidity as $Q^2$ increases, a feature not affected by our
modification to $X_0(z)$.  The effect of the new $X_0(z)$ is reflected
only in the rest of integrand in Eq.~(\ref{ff}), whose integral [{\it
i.e.}, for $V(q,z) \! \to \! 1$] is normalized to 1.  One expects that
the integrand favors larger $z$ in the original soft-wall case because
of the direct $z^3$ dependence of $X_0(z)$ than in the modified case,
since $X_0 (z) \! \to \! B_c$ as $z \! \to \! \infty$.  Since $V(q,z)$
decreases with $z$ for all $Q^2 \! > \! 0$, $F_\pi(Q^2)$ is also lower
in the original model for all values of $Q^2 \!  > \! 0$.  Our
numerical simulation indeed verifies this behavior.  The $F_\pi (Q^2)$
prediction using the modified $X_0(z)$ in Eq.~(\ref{eq:mod_vev}) is
never smaller than that of the original soft-wall model, and hence
lies at least as far from the $F_\pi (Q^2)$ data.

Second, we explore the consequences of modifying the background field
as in Eq.~(\ref{SW}), repeated here for convenience:
\begin{equation}
e^{-\Phi(z)} = \frac{e^{\lambda^2 z_0^2} - 1}{e^{\lambda^2 z_0^2} +
e^{\lambda^2 z^2} - 2} \, .
\end{equation}
As expected, the parameter $\lambda$ serves to interpolate between the
hard-wall and soft-wall models.  Specifically, for $\lambda z_0 \!
\approx \! 1$ we find that the model prediction for $F_\pi$ is very
similar to the original soft-wall result, while for $\lambda z_0 \!
\approx \! 2$ the prediction is very close to the original hard-wall
result: See Fig.~\ref{fig:formfactor}.  Beyond $\lambda z_0 \!
\approx \! 2$ the global fit does not improve much; in particular, the
$F_\pi (Q^2)$ fit is best for $\lambda z_0 \! = \! 2.1$, but at the
cost of $f_\pi \! = \! 88.0$~MeV, while at $\lambda z_0 \! =
\! 2.4$ the $F_\pi (Q^2)$ fit is not very good but one obtains $f_\pi
\! = \! 92.7$~MeV.  That results mimicking the older models are
obtained for such modest values of $\lambda z_0$ is especially
interesting because the hard-wall limit is formally recovered only
when $\lambda z_0 \! \to \! \infty$, and the soft-wall limit is
formally recovered only when $\lambda z \! \gg \! \lambda z_0, 1$.
\begin{table}[t]\vspace{-6pt}
\caption[]{Comparison of hard-wall model to the $e^{-\Phi(z)}$ of
Eq.~(\ref{SW}) with $\lambda z_0\!=\!2.1$; values in MeV (except for
$g_{\rho \pi \pi}$).}
\begin{ruledtabular}
\begin{tabular}{cccc}  
Observable & Experiment & Hard wall &$\lambda z_0 \!=\!2.1$\\
\hline
$m_\pi$            & 139.6$\pm 0.0004$ \cite{PDG}          & 139.6     & 139.6 \\
$m_\rho$           & 775.5$\pm 0.4$   \cite{PDG}          & 775.3     & 777.5 \\
$m_{a_1}$          & 1230$\pm40$      \cite{PDG}          & 1358      & 1343  \\
$f_\pi$            & 92.4$\pm0.35$    \cite{PDG}          & 92.1      & 88.0  \\
$f_\rho^{\,1/2}$   & 346.2$\pm1.4$    \cite{Donoghue}     & 329       & 325   \\
$f_{a_1}^{\,1/2}$  & 433$\pm13$     \cite{SS,Isgur:1988vm}& 463       & 474   \\
$g_{\rho\pi\pi}$   & 6.03$\pm0.07$    \cite{PDG}          & 4.48      & 4.63  \\
\end{tabular}
\end{ruledtabular}
\label{Table2}
\end{table}

As a comparison, we present results for both $\lambda z_0\!=\!1$ and
$\lambda z_0\!=\!2.1$ as well as the original soft- and hard-wall
predictions in Tables~\ref{Table1} and \ref{Table2}, respectively.
Even though the result for $\lambda z_0 \! \approx \! 2$ is very
similar to original hard-wall result, its predictions for the
positions of higher-order vector meson poles differ strongly from
those of the original hard-wall model (Table~\ref{Table3}).  The
modified $e^{-\Phi(z)}$ results exhibit a closer match to a $m_n^2 \!
\sim \! n$ vector meson Regge trajectory, in contrast to the
$m_n^2 \! \sim \! n^2$ scaling of the original hard-wall trajectory.
From Table~\ref{Table3} one finds in the hard-wall case that $m_{n+1}
\! - \! m_n$ is stable around $1010$~MeV, while in the modified
background case, $(m_{n+1}^2 \! - \!  m_n^2)^{1/2}$ is stable around
$1450$~MeV.  We conclude that the $e^{-\lambda^2\,z^2}$ tail behavior
of Eq.~(\ref{SW}) is already significant when $\lambda z_0 \!
\approx \!  2$, where the prediction for $F_\pi (Q^2)$ closely matches
the hard-wall result.  We have thus produced a model that works as
well as the hard-wall model in fitting to the low-energy data, but
that in addition reproduces the appropriate Regge trajectories
expected from linear confinement.
\begin{table}[t]\vspace{-6pt}
\caption[]{Comparison of vector meson masses and decay constants (in
MeV) for the hard-wall model and the model using $e^{-\Phi(z)}$ of
Eq.~(\ref{SW}) with $\lambda z_0 \!=\!2.1$.}
\begin{ruledtabular}
\begin{tabular}{cccc}
\multicolumn{2}{c}{Original hard wall} &
\multicolumn{2}{l}{$e^{-\Phi(z)}$ from Eq.~(\ref{SW}) with $\lambda
z_0 \!=\!2.1$} \\ \hline $m_\rho$ & $F_\rho^{1/2}$ & $m_\rho$ &
$F_\rho^{1/2}$ \\
\hline
775.6     & 329             & 777.5     & 325   \\
1780.2    & 616             & 1608.1    & 528   \\
2790.8    & 864            & 2226.8    & 611   \\
3802.8    & 1089            & 2637.5    & 644   \\
4815.2    & 1300            & 2986.6    & 683
\end{tabular}
\end{ruledtabular}
\label{Table3}
\end{table}

For completeness we also include the pion charge radius $\langle
r_\pi^2 \rangle$ results for both models.  In the modified $X_0$ model
for $B_c \! = \! 0.96$ and $A_c \! = \! 1.1$, we find $\langle r_\pi^2
\rangle \! = \! (0.485~\rm{fm})^2$.  In the modified $e^{-\Phi}$ model
for $\lambda z_0 \! = \! 1.0$, $\langle r_\pi^2 \rangle \! = \!
(0.500$ fm$)^2$, while for $\lambda z_0=2.1$ it is
$(0.576~\rm{fm})^2$.  In comparison, the original soft-wall model
gives $\langle r_\pi^2 \rangle \! = \! (0.494~\rm{fm})^2$, the
original hard-wall model gives $\langle r_\pi^2 \rangle =
(0.576~\rm{fm})^2$, and the experimental value is $\langle r_\pi^2
\rangle = [0.672(8) \ {\rm fm}]^2$~\cite{PDG}.

Having addressed the two central issues of this paper, we now discuss
an even more interesting conundrum, namely, the fact that all the fits
described here are limited in how closely they approach the full set
of $F_\pi (Q^2)$ and other low-energy data.  All of them produce
curves too shallow in $Q^2$, leading in particular to a value of
$\langle r_\pi^2 \rangle$ smaller than experiment.  We observed in
Ref.~\cite{KL} that a better fit to $F_\pi(Q^2)$ could be obtained by
loosening the fit to the other low-energy observables, particularly by
letting $f_\pi$ be somewhat smaller, a resolution that is not entirely
satisfactory.  However, as is well known~\cite{PQCD}, the functional
behavior of $F_\pi (Q^2)$ for large $Q^2$ is determined by its
partonic substructure: $F_\pi(Q^2) \propto f^2_\pi
\alpha_s(Q^2)/Q^2$.  The AdS/QCD model used here contains only
hadronic and no explicit partonic degrees of freedom, so it is perhaps
not surprising that the $F_\pi (Q^2)$ prediction has difficulty
falling as steeply as the data for larger $Q^2$.  But are partons
essential at the relatively small values ($Q^2 \! \le \!  5$~GeV$^2$)
presented in Fig.~\ref{fig:formfactor}?  If not, then we are faced
with the possibility that the pattern of chiral symmetry breaking
developed in Ref.~\cite{EKSS} and represented by the background field
$X_0(z)$ may require improvement.

\section{Conclusions} \label{concl}

In this paper we have shown, first, that the treatment in
Ref.~\cite{KL} of chiral symmetry breaking in the soft-wall model,
although not formally satisfying all the boundary conditions of that
model, nevertheless gives numerical results in agreement with those of
models possessing all the correct asymptotic behaviors.  Indeed, we
find that the soft-wall calculation of Ref.~\cite{KL} appears to
represent a best-fit limit to such improved models.

We have also developed a model that successfully interpolates between
hard-wall and soft-wall models: It reproduces the desirable numerical
features of the hard-wall model but also the Regge behavior predicted
by linear confinement that motivates the soft-wall model.  The fit to
static observables as well as $F_\pi (Q^2)$ is as good as that from
the hard-wall model, but the excited $\rho$ masses now follow the
desired pattern $m_n^2 \! \sim \! n$.

Nevertheless, all of the models proposed here and in Ref.~\cite{KL}
share the properties that their predictions for $F_\pi (Q^2)$ are too
shallow for larger $Q^2$ values, unless one relaxes the fit to one or
more of the static observables (particularly $f_\pi$).  We have argued
that the most likely culprits for this discrepancy are either the
application of a purely hadronic AdS/QCD model to a region (large
$Q^2$) where partonic degrees of freedom become appreciable, or the
need for an improvement to the pattern of chiral symmetry breaking
introduced via background fields.  Both directions are well worthy of
further research.

\appendix*
\section{Numerical Procedure}
Solutions to Eqs.~(\ref{eqVAdS})--(\ref{Xeqn}) are obtained
numerically using subroutines provided by {\it Numerical Recipes\/}
(NR)~\cite{NR}, and all terminology provided in this appendix is
discussed in that well-known standard reference.  These differential
equations are solved using the shooting method because the boundary
conditions are defined on both ends of the integration range,
generically labeled [$x_1, x_2$].  This procedure consists of several
steps, which we now outline.  Starting at the initial point $x_1$ of
the integration, suppose that $N$ boundary conditions are to be
specified but only $n_1$ are provided.  This leaves $n_2 \! \equiv
\!  N \! - \! n_1$ freely specifiable starting points, from which one
can form an $n_2$-dimensional vector {\bf V}, the initial value of
which is the first guess for the remaining boundary conditions.  We
then use the combined boundary conditions to integrate the
differential equations and obtain $y(x_2)$ ($y$ being the generic
dependent variable label) at the endpoint $x_2$ of the integration.
Here one forms an $n_2$-dimensional discrepancy vector {\bf F}, which
is the difference between the result $y(x_2)$ of the numerical
integration and the specified boundary conditions at $x_2$.  The
integration is performed via an adaptive stepsize Runge-Kutta (RK)
method with constants provided by Cash and Karp.  The solution is then
obtained by optimizing the vector {\bf V} such that the vector {\bf F}
converges to zero, by means of the globally-convergent Newton-Raphson
method.

As with any numerical procedure, this calculation has an inherent
rounding error.  There are two main sources for these errors, the RK
integration and the Newton-Raphson method.
\begin{enumerate}
\item The RK integration routine naturally provides in its result a
5th-order truncation error (called $\frac{1}{15}\Delta$ in NR) on the
dependent variable.  This truncation error scaled by some combination
of the function $y$ and its derivative $dydx$ (using the terminology
of Ref.~\cite{NR}) evaluated at the initial point of each step in the
integration determines the size $h$ of the next step, but more
importantly, drives the desired accuracy of our calculation.  In
particular, the truncation error is obtained by taking the difference
between the result calculated by adaptive stepsize RK (at 5th-order)
and the result calculated by classical 4th-order RK.  The small
dimensionless parameter $eps$ (fractional error) correlated to the
truncation error is then defined by $\Delta \! = \! eps \! \times \!
yscal(i)$, where $i$ is the index of the particular differential
equation among those simultaneously solved that contributes the most
to the truncation error (the ``worst offender'').  We choose the
particular recipe for $yscal$ in our calculation to be $yscal(i) \!  =
\! |y(i)| \! + \! |h \! \times \!  dydx(i)|$ (see NR for other
recipes).  Lastly, we use $eps \!  = \! 10^{-6}$ in our calculation.
\item The Newton-Raphson method introduces several small parameters to
assure convergence:
\begin{itemize}
\item The small parameters ALF and TOLMIN are introduced to guarantee
the proper convergence of the Newton-Raphson subroutine and therefore
affect the final result of the calculation only indirectly.  ALF
functions as an indicator that the evaluated function has ceased to
decrease appreciably, as determined by the Newton-Raphson method.  In
this case the subroutine backtracks along the Newton direction until
an acceptable rate of decrease is obtained.  TOLMIN sets the criterion
for identifying spurious convergence, in which case the subroutine
warns of the event and requires the introduction of a new initial
point.  These parameters are chosen to be ALF = $10^{-4}$ and TOLMIN =
$10^{-6}$.
\item TOLX sets the convergence criterion for $\delta x$, which
represents $\delta${\bf V} in the shooting method algorithm.  Its
convergence criterion is set by TOLX = $10^{-7}$.
\item TOLF sets the convergence criterion on {\bf F} values; TOLF =
$10^{-4}$.
\end{itemize}
Further details can be found in Ref.~\cite{NR}.
\end{enumerate}
All of these parameters are at least as small as (in most cases,
appreciably smaller than) the measured uncertainties of the low-energy
QCD observables to which we fit our models.  As such, we can safely
assume all the errors in our fits come primarily from experimental
uncertainties that have been provided in the text, and we do not
include further error analysis in our paper.  Indeed, the chief figure
of merit is the closeness of fit to the central values of certain
hadronic quantities, as we now discuss.

The meson masses $m_{\rho}$, $m_{\pi}$, and $m_{a_1}$ are identified
as poles in the solutions to Eqs.~(\ref{eqVAdS})--(\ref{Xeqn}) [see,
{\it e.g.}, Eq.~(\ref{spectral})].  One identifies pole masses by
noting that the numerical solution near $q^2 \! \to \!  m^2$ crosses a
singularity and flips sign.  Once a mass eigenvalue is identified, its
individual bulk-to-boundary propagator is the solution to the relevant
wave equation at that value of $q^2 \! = \! m^2$, using the shooting
method just described.  With the normalization for $g_5$ chosen,
solutions to Eqs.~(\ref{eqVAdS})--(\ref{Xeqn}) at $q^2 \! = \! 0$ are
used to derive the decay constants $f_{\rho}$, $f_{\pi}$, and
$f_{a_1}$ [as in, {\it e.g.}, Eq.~(\ref{fpi})].  The overall quality
of our fits is driven chiefly by (the central values of) the three
best-measured meson quantities, $m_\rho$, $m_\pi$, and $f_\pi$.  The
other parameters in our models such as $A_c$, $B_c$, and $\lambda$ are
then allowed to float freely.

\section*{Acknowledgments}
This work was supported by the NSF under Grant No.\ PHY-0456520.

\end{document}